\documentclass[samsmath,amssymb,floatfix,superscriptaddress,aps]{revtex4}
\usepackage{graphicx}
\begin{document}
%\draftMos
%\twocolumn[
%\hsize\textwidth\columnwidth\hsize\csname @twocolumnfalse\endcsname
%\draft
\title{Colloidal systems with competing interactions: \\ from an arrested
repulsive cluster phase to a gel}
\author{Juan Carlos Fernandez Toledano} 
\affiliation{ {Dipartimento di Fisica and  INFM-CRS-SOFT, Universit\`a di Roma {\em La Sapienza}, P.le A. Moro 2, 00185 Roma, Italy} }
\affiliation{Grupo de F\'isica de Fluidos y Biocoloides, Departamento de F\'isica Aplicada, Facultad de Ciencias, Campus Fuentenueva S/N, 18071 Granada, Spain }
\author{Francesco Sciortino} \affiliation{ {Dipartimento di Fisica and
  INFM-CRS-SOFT, Universit\`a di Roma {\em La Sapienza}, P.le A. Moro
  2, 00185 Roma, Italy} }
%\author{Roque Hidalgo-Alvarez}
%\affiliation{Grupo de F\'isica de Fluidos y Biocoloides, Departamento de F\'isica Aplicada, Facultad de Ciencias, Campus Fuentenueva S/N, 18071 Granada, Spain }
\author{Emanuela Zaccarelli} \affiliation{ {Dipartimento di Fisica and INFM-CRS-SOFT, Universit\`a di Roma {\em La Sapienza}, P.le A. Moro 2, 00185 Roma, Italy} }
        
\begin{abstract}
We report an extensive numerical study of a charged colloidal system
with competing short-range depletion attraction and long-range
electrostatic repulsion. By analizing the cluster properties, we
identify two distinct regions in the phase diagram: a state composed
of stable finite-size clusters, whose relative interactions are
dominated by long-range repulsion, and a percolating network. Both
states are found to dynamically arrest at low temperatures, providing
evidence of the existence of two distinct non-ergodic states in these
systems: a Wigner glass of clusters and a gel.
\end{abstract}

\maketitle

\section{Introduction}
The origin of low-density non-ergodic states in colloidal systems is a
matter of continuous debate and ongoing
research\cite{Poon98a,Tra04a,Cip05a,Zac07a,Pue08a}. Several different
mechanisms may concur in the formation of these arrested states,
depending on the relative ratio between the thermal and binding energy
and on the shape and symmetry of the interaction potential.  When
particles interact via excluded volume interactions complemented by a
spherically symmetric, attractive potential, it has been shown that
the formation of a gel structure takes place concurrently with a
spinodal decomposition process. Gelation results from an arrested
phase separation\cite{Lu08a}.
%Different scenarios arise when interaction is no longer spherically symmetric, but acquires directionality or patchy properties, since in this case the phase separation process can be confined in a small region of density (depending on particle valence)\cite{Zaccagel,Bia06a} and ideal network formation in equilibrium can be approached\cite{Zac07a,genova}.
%To approach ideal network formation at arbitrarily low densities, 
%the use of non-spherical `patchy' interactions is needed. 
A different scenario occurs when colloidal particles have a residual
electrostatic charge which builds up an additional long-range
repulsion in the effective colloid-colloid interaction. This term is
often modeled as a Yukawa potential with Debye screening length $\xi$
to take into account the presence of the solvent and
counterions\cite{Lik01b}. In apolar solutions or under low salt
conditions, when particles are sub-micron sized, $\xi$ can become
comparable to the particle dimension. This long-range repulsive term
can coexist with a short-range attraction (which can be induced for
example via depletion interactions), generating a competition between
aggregation driven by the attractive part of the potential and the
stabilizing role of the repulsion, which may ultimately suppress the
macroscopic phase separation.
%allow for the exploration in a quasi-equilibrium approach of the phase space corresponding to large attraction strengths without encountering an intervening phase separation\cite{Zac07a}. 
Indeed, it has been shown that the addition of a long-range repulsion
of moderate strength can shift to larger attraction strengths (or
lower temperatures) the phase separation\cite{Pini00,Cha07b},
eventually inhibiting it\cite{Arc07a,Tarzia,Ort07a,Ort08a}.  In this
case a microphase separation into
clusters\cite{Strad04,Imp04,Sedg04,Bagl04,Card06} of a preferred
cluster size and shape takes place, depending on repulsion
parameters\cite{Mos04a}.  When repulsion is moderately short-ranged,
i.e. $\xi/\sigma \lesssim 0.5$ with $\sigma$ being the diameter of the
colloidal particle, it was observed both in experiments and in
simulations that elongated clusters are formed at low enough
temperature $T$ \cite{Bartlett04,Sciobartlett}. The repulsion between
such clusters is relatively weak so that they tend to form at low $T$
quasi-ordered columnar structures\cite{WuCaoPhysA,Candia}.  At large
enough packing fraction,
%(i.e. $\phi \gtrsim 0.125$), 
the clusters are found to merge into a percolating network
\cite{Bartlett04,Sciobartlett}. This network of clusters exists
at low enough $T$ and undergoes dynamical arrest, so that a gel state
can be properly identified\cite{Zac07a}.

For cases where the repulsion term is considerably longer-ranged,
i.e. $\xi/\sigma \gtrsim 1$, simulations at low enough colloidal
densities\cite{Sci04a} have reported the presence of a Wigner glass of
clusters. This corresponds to a disordered state of polydisperse
clusters (due to finite $T$) which do not percolate and are actually
arrested due to the long-range repulsion, in analogy with the Wigner
glass reported by Chaikin and coworkers\cite{Cha1982,Cha1989} for
charged colloidal particles under very dilute conditions, stabilized
by the Coulomb repulsion.  Recently, a comparison between theory and
simulations\cite{Zac08a} of Yukawa particles has shown that the ideal
Mode Coupling Theory (MCT) provides a quite accurate description of
the formation of a particle Wigner glass.  The MCT predictions for
Yukawa particles have also been exploited for interpreting arrest into
a Wigner glasses of clusters, in systems with competing
interactions\cite{Sci04a,ChenPRE}.  Indeed, once clusters are assumed
spherical and monodisperse (in size), the effective cluster-cluster
interactions can be modeled in terms of a Yukawa potential, with the
same screening length as the one acting between single particles but
with a renormalized amplitude\cite{Sci04a}.

These earlier works call for additional investigations, in order to
further question the existence and the stability of a Wigner glass of
clusters, as well as a deeper understanding of cluster-cluster
interactions.  To this end, it is also relevant to mention a recent
simulation study \cite{Cha07a} where clusters were observed to arrest,
at not-too-low density, by percolation rather than by
repulsion. However, differences in the simulation protocol of this work are present with
respect to that used in \cite{Sci04a}, in particular history
of quench and quench rate, as well as a shorter cut-off distance for
the long-range repulsion potential. Hence, a more
comprehensive study of these models in a wide region of packing
fraction $\phi$ and temperature $T$, fully accounting for the long
range nature of the repulsive interactions, is needed.

In this work we report an extensive simulation study which aims 
at elucidating in detail the phase diagram of colloidal systems interacting
with both short-range attraction and long-range repulsion,
encompassing states well below, at the crossing and well above the
percolation line at low enough $T$.  We characterize in detail the
properties of the aggregates that are formed, both when they exist in
finite size objects (clusters) and when they merge into a percolating
network. We analyze both particle-particle and cluster-cluster
correlations to show that, at low enough packing fractions, the system
self-organizes into stable, long-living clusters which do not
percolate and do not form ordered structures. The interactions between
these clusters can be characterized, for low and intermediate $\phi$,
in terms of renormalized long-range repulsive interactions of Yukawa
form, as previously hypothesized\cite{Sci04a}. Most importantly,
despite the high polydispersity and non-sphericity of the clusters, we
find that the screening length of cluster-cluster interactions remains
unchanged with respect to that of particle-particle interactions,
while the repulsion amplitude is found to increase with particle
density.  At larger $\phi$, when the shape of the clusters starts to
significantly deviate from the spherical one and increased packing
leads to the arising of branching events, a crossover takes
place, ultimately leading to the percolation of the clusters associated to a
gel transition.  We also monitor the dynamics of the two regions,
calling for the existence of two distinct non-ergodic states in this
low-density, low-$T$ part of the phase diagram.

\section{Description of the simulation protocol and analysis}
We consider a system of $N=1000$ particles of unit mass $m$
interacting with a total pair potential composed of a short-ranged
attractive part, modeled for convenience as a generalized
Lennard-Jones potential\cite{Vli99} with exponent $\alpha=100$, and of
a long-range screened electrostatic repulsion, modeled as a Yukawa
term,
\begin{equation}
V(r)=4\epsilon\left[\left(\frac{\sigma}{r}\right)^{2\alpha}-\left(\frac{\sigma}{r}\right)^{\alpha}\right]+
A \frac{\exp{-r/\xi}}{r/\xi}
\end{equation}
where $A$ is the amplitude of repulsion and $\epsilon$, the depth of
the attractive part, is chosen as the energy unit.  Time is measured
in units of $\sqrt{m \sigma^2 / \epsilon}$. The choice of $\alpha=100$,
also studied in \cite{Sci04a, Mos04a}, ensures a very short-range
attraction, corresponding to bond formation only within the first neighbour 
shell. According to the extended law of corresponding states for
spherical short-ranged attractions\cite{Nor00aJCP}, such choice is
generic for any width and shape of the attractive
potential\cite{Lu08a}. Note that also \ the cluster ground-state properties
have been shown to be invariant for $\alpha \geq 18$\cite{Mos04a}.  We
fix the parameters of the Yukawa potential to $\xi=2\sigma$ and
$A=0.2\epsilon$. For such values a microphase separation into clusters
exists\cite{Sci04a, Mos04a}. To correctly take into account the
long-range nature of the interactions we solve the equations of motion
using Ewald summation\cite{ewald}. Indeed, we have compared results
obtained in this way with those based on the use of a finite (although
large) cut-off.  For our choice of parameters, an underestimate of
about $10\%$ for the repulsive potential energy is provided by the use
of a finite cutoff at $8\xi$, in agreement with previous studies
\cite{Gia05a}. Hence, despite the significantly increased
computational cost, it appears crucial for the case under study to
treat the long-range repulsive term with Ewald sum, in order to
discriminate cases where clusters truly form a disconnected or a
percolating state and to address satisfactorily the nature of the
arrested state(s) for the chosen value of $\xi$.

To model the motion of colloidal particles in a solvent, we use
Brownian dynamics simulations with time step $\delta t=0.005$ and bare
diffusion coefficient $D_0=0.005$.  For this choice, the crossover
from ballistic to diffusive regime, for isolated particles, takes
place for $t\sim 10$.  We do not treat explicitly the effect of the
solvent, i.e. hydrodynamic interactions are neglected, but we do not
expect these to provide significant changes to the long-time
structures that we observe \cite{Yam08a}.  We neglect any change in
the electrostatic parameters, in particular of $\xi$, with increasing
colloid packing fraction $\phi$. For particles of the order of $\mu
m$, this variation is expected not to be significant with respect to
the particle diameter\cite{2003JPCM.Royall}.

The system was initially prepared at several densities and high $T$,
and later it was slowly equilibrated to successively lower $T$.  Our
aim is to study the approach (from the equilibrium side) to a dynamic
arrest transition, rather than a rapid quench inside the region where
arrest is observed. We study 9 isochores and several $T$ in the
low-$\phi$ region up to $\phi=0.20$.  Equilibration was carried out in
a NVT ensemble, followed by production runs for data collection and
analysis.  For very low $T$ (depending on the studied $\phi$), a true
equilibration is no longer possible. Monitoring for example the
energy per particle, the system at first shows a robust decrease
towards an apparent equilibrium state, but then the system starts to
display a very slow (logarithmic in time) energy drift, typical of an
approach to dynamic arrest and trapping in a metastable state.
% A typical evolution with $T$ of the energy per particle for a single isochore is shown in Fig.~\ref{fig:energy}.

The connectivity properties of the system have been monitored by
studying the interparticle bonding.  Two particles are considered
bonded when the distance between them is smaller than the position of
the local maximum in the interaction potential, i.e. $r<
r_b=1.072\sigma$.  When a state point has reached equilibrium, we
collect several independent realizations of the system and calculate
the distribution $n(s)$ of clusters of size $s$ over time. Moreover,
we check whether the largest cluster that we find in the configuration
has spanned the whole box at least in one direction. If at least
$50\%$ of the considered realizations contains a spanning cluster, the
state point is a classified as percolating. For state points which do
not percolate at the studied $T$, we monitor the cluster properties in
time, finding that the clusters have a finite, but large lifetime.  If
the cluster size distribution shows a maximum for a finite value of
$s$, indicating that a preferred size for the clusters exists, we
classify this state point in the cluster phase region.

\section{Finite Size Clusters and Percolation: disordered, metastable states}
\begin{figure}[tbh]
\centering
\includegraphics[width=0.6\textwidth]{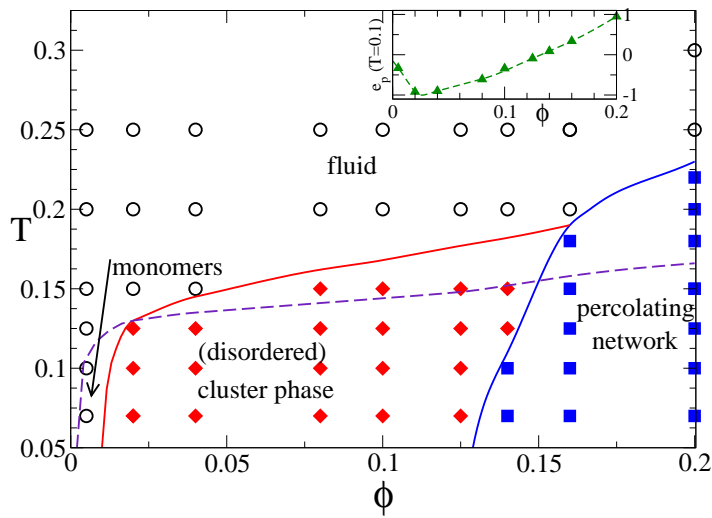}
\caption{Phase diagram of the studied system. Open symbols indicate
  fluid states (no cluster phase and non-percolating).  Filled
  diamonds indicate points within the cluster phase region
  (i.e. showing a maximum at finite $s$ for $n(s)$) and filled squares
  refer to percolating state-points. Full lines, defining the cluster
  region and the percolation region, and the dashed line, defining the
  region (i.e. the state points below the line) where the energy starts to
  display a slow logarithmic drift, are drawn as guides to the eye. At
  low enough $\phi$ (i.e. $\phi \lesssim 0.02$) the cluster size
  distribution is peaked at $s=1$. Inset: potential energy per
  particle $e_p$ versus packing fraction. A minimum is found at very
  low $\phi$; $e_p$ becomes positive for $\phi >0.125$, i.e. close to
  the location of the percolation transition.}
\label{fig:phase}
\end{figure}

In Fig.~\ref{fig:phase} we report the phase diagram of the system in
the studied $(\phi, T)$ region, highlighting the loci of points where
(i) a cluster phase is observed and (ii) the system percolates.  We
also report a (dashed) line delimiting the region where a true
equilibration cannot be reached, meaning that, for all investigated
state points below the line, the potential energy keeps displaying a
very slow, logarithmic drift for the entire duration of the
simulation.  This region largely belongs to the portion of the phase
diagram where stable clusters and percolation are present.

We notice that in the absence of repulsion, the critical temperature
for the short-ranged attractive potential studied here is found at
$T_c\simeq 0.24$\cite{Sci04a}, while the critical packing fraction is
found to be approximately $\phi_c \simeq 0.27$, following the behavior
for spherical attractive potentials in the short-range
limit\cite{Lar08a}. Hence, the addition of a long-range repulsion
suppresses the tendency to phase separate macroscopically, at least in the
$T$-window that we have studied, i.e. for $T\geq0.05$.  It becomes
thus possible to reach lower temperatures in one-phase condition. At
these low $T$, the lifetime of the inter-particle bond increases
significantly and clusters behave as effective long-lived aggregates.

Fig.~\ref{fig:phase} shows that for $\phi\leq 0.125$ the system
never percolates at all studied $T$.  For very small $\phi$
(e.g. $\phi=0.005$) the system organizes into very small clusters, and
the maximum of $n(s)$ is located to $s=1$ at all studied $T$.  We can
consider this very low-$\phi$ region to be essentially in a
`monomeric' state. As $\phi$ is increased, a maximum in the cluster
distribution $n(s)$ develops for $T\lesssim 0.15$, and we can identify
a stable cluster phase, according to our definition
discussed above.  In this region, clusters can be monitored in time
and we observe that they continuously exchange particles with each
other, breaking and reforming bonds, while on average existing at all
times in different realizations. For $T\lesssim 0.1$, clusters become
more and more long-lived, since the time necessary to break a bond
becomes comparable to the simulation time. Similarly, the encounter
between two different clusters is quite rare, due to the presence of
the long-range repulsion. Essentially, clusters become
frozen. Finally, for $\phi \gtrsim 0.14$ we observe percolating
states.  Hence, at low $T$ or equivalently for high attraction
strengths, the system displays a transition from monomers to a stable
finite-cluster phase and eventually to a particle network.  These
findings are in very good agreement with recent confocal microscopy
experiments carried out for a charged colloidal suspension with
depletion interactions under low screening
conditions\cite{royallaachen}, consistent with the case discussed in
the present work. 

In the inset of Fig.~\ref{fig:phase}, we report the dependence of the potential energy per particle $e_p$,  on $\phi$ for
$T=0.1$. Data are collected at the end of the long simulation runs (due to the presence of the slow logarithmic aging). 
We notice that the energy has a minimum for $0.02 < \phi <
0.04$ and then becomes positive for $\phi > 0.125$, i.e. in the region
where percolation is observed. This points to the fact that only at
very low $\phi$ particle clustering is dominated by the attractive interactions,
while the interplay with repulsion becomes important already for $\phi \gtrsim 0.04$.
The increase in the energy arises from  the
large number of neighbours within the  repulsive range of the interaction, which overcomes
the gain associated to bonding.

\begin{figure}[tbh]
\centering \includegraphics[width=0.8\textwidth, angle=0]{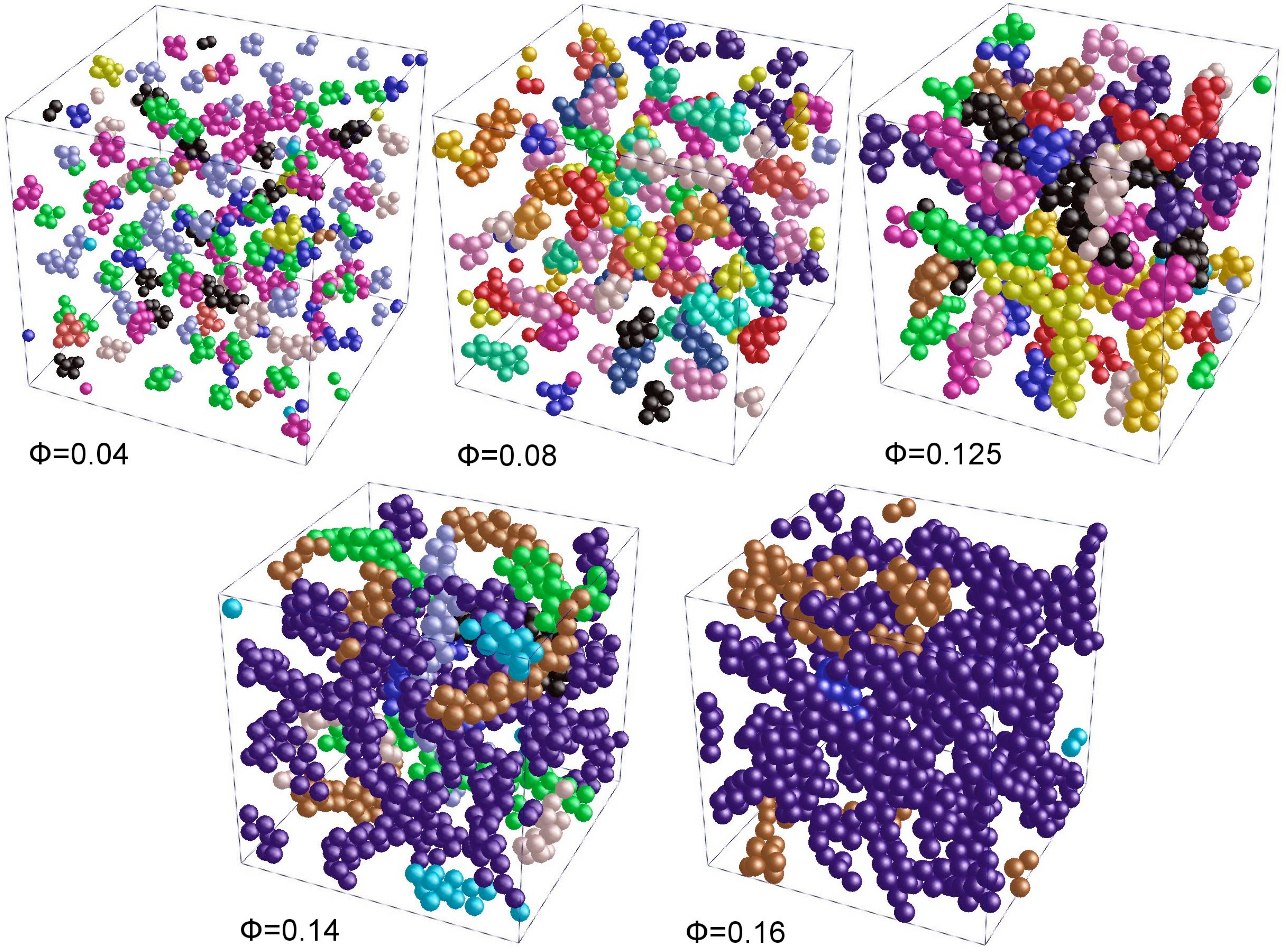}
\caption{Snaphots of the simulated system for serveral different $\phi$ at $T=0.10$.
Different clusters are drawn in different colors according to
  particle number in the cluster. For $\phi \leq 0.125$ (top row), the system
  forms at all times several distinct clusters which are not connected
  and do not span the simulation box; while for $\phi \geq 0.14$ (bottom row) a
  percolating cluster exists at all times.  }
\label{fig:clust}
\end{figure}

To visualize the cluster formation and to highlight the fact that the
system at low $\phi$ neither percolates, nor forms ordered
structures, we report in Fig.~\ref{fig:clust} some snaphots of the
system for different $\phi$ along the isotherm $T=0.10$. Different
clusters are drawn in different colors, according to particle
number. Although exchange of particles between clusters as well as
cluster branching and breaking processes are sometimes observed, a
similar picture of clusters to that reported in Fig.~\ref{fig:clust}
exists at all times. The emergence of a percolating cluster arises,
for the reported temperature, at $\phi=0.14$, and rapidly involves the
majority of particles as $\phi$ grows further.  We notice that in the
cluster phase region, clusters are polydisperse, both in size and in
shape. While at low $\phi$ they are always rather spherical, they tend
to become more and more elongated with increasing packing
fraction. This phenomenon was observed in previous
simulation\cite{Sciobartlett} and experimental\cite{Bartlett04}
studies for a system with a much shorter screening length.  A study of
the ground state properties of isolated clusters\cite{Mos04a} has
shown that, while in the short screening length case quasi
one-dimensional cluster growth is energetically favoured, giving rise
to the peculiar Bernal spirals\cite{Bartlett04,Sciobartlett}, for the
present study the expected ground state cluster structure is much more
spherical, although with some degree of anisotropy. Comparing the
ground state structure of clusters of a certain size, shown in Fig. 5
of Ref.\cite{Mos04a} for the two cases, it is evident that the average
number of nearest neighbours is dramatically different: while in the
Bernal spiral, particles have always exactly 6 neighbours (since there
is no difference between bulk and surface), in our case particles in
the interior of the cluster have a coordination close to 12
neighbours, while those on the surface a much smaller one (close to
6). Hence, the resulting average coordination is close to 8. 

\begin{figure}[tbh]
\centering \includegraphics[width=0.6\textwidth]{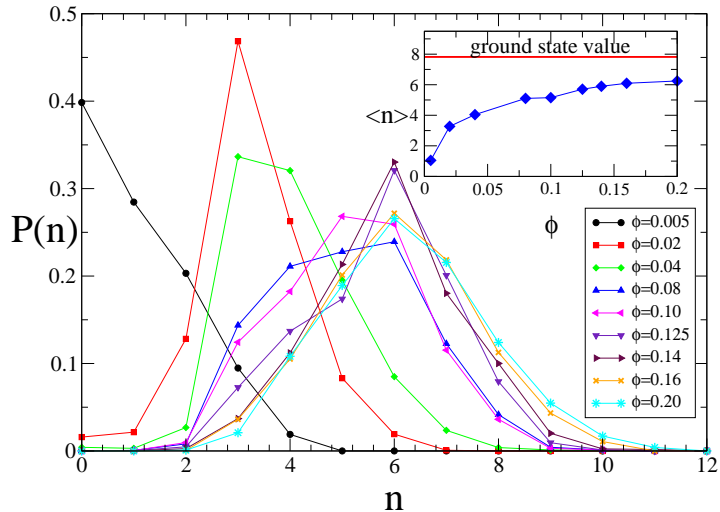}
\caption{Distribution of nearest neighbours $P(n)$ for all studied
  $\phi$ and $T=0.1$. Inset: dependence on $\phi$ of the average
  number of neighbours $<n>$ and comparison with isolated clusters ground state
  predictions from \protect\cite{Mos04a}.}
\label{fig:nn}
\end{figure}

To gain a better understanding of the cluster shape and local
structure for the present case, we report in Fig.~\ref{fig:nn} the
(average) distribution of nearest neighbours $P(n)$ for all studied
$\phi$ for $T=0.1$, and the average number of neighbours $<n>$ in the
associated inset, which is compared to the ground state predictions.
While, as expected, the number of nearest neighbours at first
increases with $\phi$, above percolation it roughly stops evolving, so
that $<n>$ does not grow much above 6. Indeed, the distribution tends
to saturate and to remain always peaked around 6 neighbours, despite a
gradual increase in the number of particles with large number of
neighbours. Monitoring the evolution of $<n>$ with $T$ (not shown), we
do not observe the presence of non-monotonic effects, which were found
in the low-screening length case\cite{Sciobartlett}, suggesting that
the long-range repulsion never facilitates the formation of compact
structures.  Comparing with the isolated cluster study of
Ref.\cite{Mos04a}, we can argue that the interactions between clusters
act against the compaction of the clusters and favor the formation of
elongated structures.  Indeed, if we imagine cluster-cluster
interactions to be ruled by the long-range repulsion (as we will prove
below), clusters will occupy much more efficiently the space and, at
the same time, reduce the total potential energy due to increased
average cluster-cluster distance by growing in an elongated rather
than spherical manner.

\begin{figure}[tbh]
\includegraphics[width=0.6\textwidth]{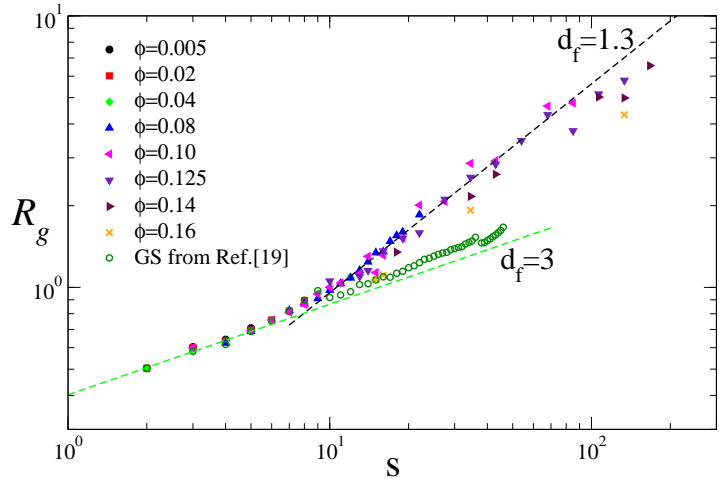}
\caption{(a) Radius of gyration of finite-size clusters for
  all studied $\phi$ at $T=0.1$. Also shown are the ground state
  calculations (open circles) for isolated clusters taken from \protect\cite{Mos04a}. Dashed
  lines indicate power-law behavior $s^{1/d_f}$, namely spherical ($d_f=3$) for
  small sizes and quasi-linear ($d_f=1.3$) for large sizes.}
\label{fig:gyr}
\end{figure}

Next, we examine the behaviour of the radius of gyration of clusters
of size $s$, defined as $R_g(s)\equiv\frac{1}{s^{1/2}} \langle
\left[\sum_{i=1}^s ({\bf r}_i -{\bf R}_{CM})^2\right]^{1/2} \rangle $,
where ${\bf r}_i $ are the coordinates of particle $i$, ${\bf R}_{CM}$
is the cluster center of mass and the average is performed over all
particles of size $s$.  We report $R_g(s)$ in Fig.~\ref{fig:gyr} for
all studied $\phi$ at $T=0.1$.  Percolating clusters are not included
in the analysis.

We notice that, in agreement with data in Ref.\cite{Sciobartlett}, the
dependence of $R_g$ on cluster size below the percolation threshold
does not show a dependence on $\phi$, although larger $\phi$ values allow
sampling of larger cluster sizes.  However, as $\phi$ grows above
percolation, clusters gradually become slightly more compact,
displaying a smaller $R_g$ at comparable sizes. We can try to identify
a growth law for $R_g$ with size.  For $s \lesssim 10$, clusters are
compact objects, as expected. Defining a fractal dimension for
clusters $d_f$ as the power law exponent of $R_g \sim s^{1/d_f}$, we
find that for $s \gtrsim 10$, clusters lose their compactness and a
quasi-linear growth is observed, roughly compatible with an exponent
$1.3$.  This is close to what observed for the Bernal spiral case. The
figure also shows the corresponding data for the ground state
configuration of isolated clusters, taken from Ref.\cite{Mos04a}.  In
comparison with the isolated clusters ground states, the clusters
observed in the simulations are significantly less compact, suggesting
that interactions between clusters and the non-negligible role of
entropy induce a one-dimensional growth already at smaller sizes.
%However, for $\phi$ above percolation the data seem to approach the ground state calculations at least on intermediate scales (i.e. $s\sim 20$), while they still recover a fastergrowth for larger sizes. 
Hence, the system always remains very far from isolated clusters
ground state predictions. Probably this results from a combination of
effects: on one hand, the increased bond and cluster lifetime does not
allow an effective restructuring of the clusters towards their
preferred configuration, on the other hand entropy and cluster-cluster
interactions act against spherical growth.  In summary, we find the
system existing in highly disordered, metastable states, both above
and below percolation. This is very clear looking at the Movies,
attached for several studied $\phi$ at $T=0.1$ as supporting material
to the present work\cite{movies}.  These results point to the
existence of two distinct metastable disordered states, that we are
tempted to identify as a Wigner glass of clusters below percolation
and a gel state above percolation. To verify whether these two states
are kinetically arrested, we will examine the dynamics of the system
in the following. However, we can already report that an ordered
low-$T$ state is never reached during the course of our simulations at
any studied $\phi$, despite the slow equilibration path that we have
used, probably due to the large screening length used. Indeed, a
recent study\cite{Candia} where a much shorter screening length was
used (although the potential studied there was not Yukawa-like but
exponential) reported the formation of a columnar phase during the
course of the simulation: the tendency to order was increased in that
case by the reduced repulsive barrier, facilitating formation and
breaking of bonds.

\begin{figure}[tbh]
\includegraphics[width=0.6\textwidth]{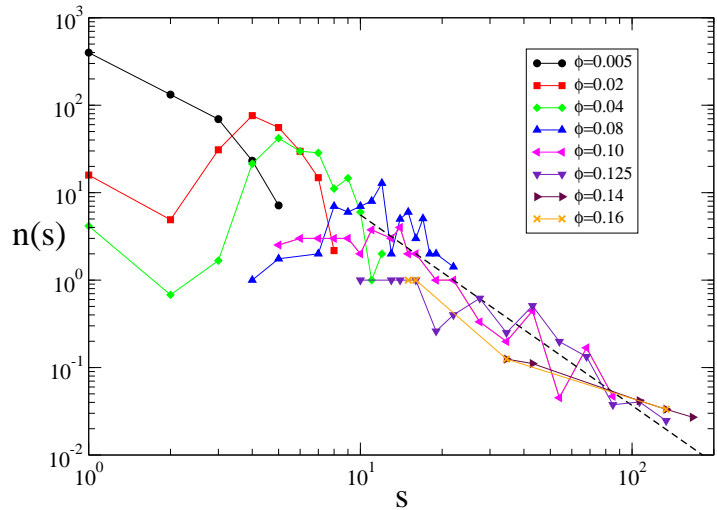}
\caption{
Cluster size distribution for all studied $\phi$ at $T=0.1$;
 only non-percolating clusters are included. The dashed line is the
 prediction for random percolation $n(s) \sim s^{-2.2}$.}
\label{fig:clustdist}
\end{figure}

Finally, we report the cluster size distribution along the same
isotherm, for all studied $\phi$, in Figure \ref{fig:clustdist}.  We
observe a number of relevant phenomena: (i) the emergence of a clear
peak between $0.02 \lesssim \phi \lesssim 0.125$, delimiting the
cluster phase region for this $T$ as we have defined it and whose
boundaries have been represented in Fig.~\ref{fig:phase}. The peak
arises at successively larger sizes as expected; (ii) the gradual
disappearance of monomers ($s=1$), which from dominant at low $\phi$,
in what we have called the monomeric phase, become absent. Not only
monomers disappear, but also small clusters of increasing sizes
gradually disappear, and the system exists in the form of finite-size
clusters; (iii) for increasing $\phi$, numerical noise becomes
important, due to the small number of finite clusters, but, within
such noise, the cluster distribution is consistent with the random
percolation power-law prediction, i.e. $n(s) \sim s^{-2.2}$.

We conclude this paragraph by observing that, in the present study, we
do not find any signature of reentrant percolation as it was found in
the short-range repulsion case \cite{Sciobartlett}. In the latter
case, at high enough $\phi$, percolation was observed at first for
high $T$, due to random aggregation of particles, then a restructuring
into the preferred shape was observed, giving rise to a
non-percolating regime for intermediate temperatures, and finally to a
new random percolation of the spiral-like clusters at low $T$. Here,
in the investigated region (up to $\phi=0.20$), we do not observe such
behavior.

\section{A more careful look at the clusters: Intra-cluster and Inter-cluster properties}

We have seen so far that, at low $\phi$, the system remains organized
into several clusters for all studied temperatures. We have tried also
deeper quenches (e.g. $T=0.01$), where the system remains
far-from-equilibrium, and we never observe the coalescence of such
clusters due to the long-range repulsion.  We have also
seen that the clusters are polydisperse, even though a preferential
size emerges as a peak in the $n(s)$, and that no long-range order is
present, i.e.  we do not observe the presence of any columnar or
lamellar phase.  However we notice that, at low $T$, a certain degree
of order develops inside the clusters.

\begin{figure}[tbh]
\includegraphics[width=0.475\textwidth]{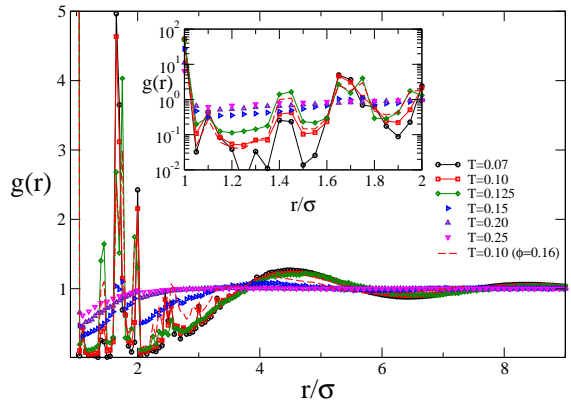}
\caption{Radial distribution function $g(r)$ for particle-particle
  correlations along the isochore $\phi=0.08$ ( within the cluster
  phase region) and several studied $T$. Also, results for $\phi=0.16$
  (within the percolating region) at low $T$ (dashed line) are
  reported to show the invariance of the locally preferred positions
  and liquid-like disorder at large distances in the two connective regimes. to
  Inset: magnification highlighting the contact peak.}
\label{fig:gr}
\end{figure}

By looking at the particle-particle radial distribution function
$g(r)$, shown in Fig.~\ref{fig:gr} for $\phi=0.08$, clear sharp peaks
arise with decreasing $T$. After a very large maximum at contact
(highlighted in the inset), enhanced correlations are found in
correspondence of specific discrete distances indicating locally
preferred geometries (e.g. triangular, tetrahedral, linear order,
etc.). However, liquid-like disorder is retained after the 2nd peak of
the $g(r)$. This behaviour is invariant whether we consider states
below and above percolation, as also shown in the figure.  A parallel
analysis of the structure factor (not shown) confirms these
observations. This indicates that, at low $T$, particles inside the
clusters tend to occupy preferential ordered positions, thus providing
quasi-crystalline character to the inside of the clusters. However,
this order is lost already after the second neighbour shell, both in
the cluster region and in the percolating one.

\begin{figure}[tbh]
\includegraphics[width=0.6\textwidth]{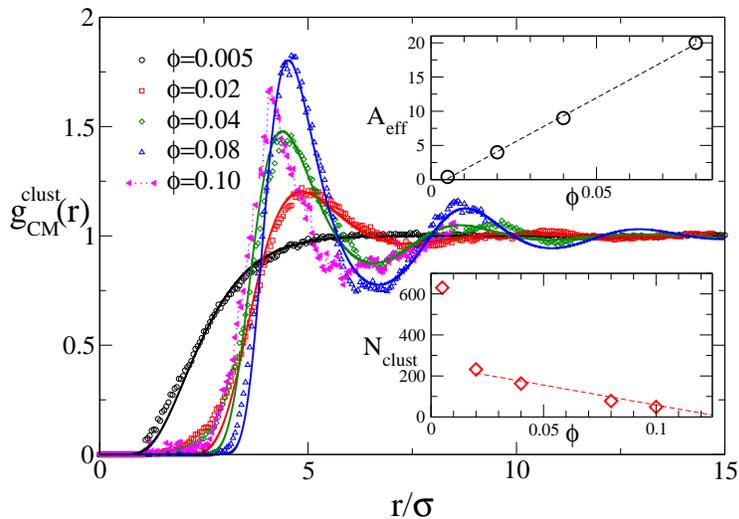}
\caption{Radial distribution function between clusters centers of mass
  for $T=0.10$ and various studied $\phi$ in the cluster phase region.
  Symbols are simulations results, lines are theoretical calculations
  based on HNC solution of the OZ equation for a renormalized Yukawa
  interactions. Insets: (a) renormalized amplitude $A_{eff}$ of the
  Yukawa cluster-cluster interactions and (b) clusters number density
  (in units of $\sigma^{-3}$) dependence on $\phi$.  Dashed lines are
  linear fits.}
\label{fig:grCM}
\end{figure}

After the analysis of particle-particle correlations, it is
interesting to consider cluster-cluster correlations to get an idea of
what mechanisms regulate the interactions between different clusters
and why they do not percolate at sufficiently low $\phi$. In a
previous work\cite{Sci04a}, it was hyphothesized that cluster
interactions may be of renormalized Yukawa form, maintaining the same
screening length as the underlying particle-particle interactions and with an
increased repulsion strength with increasing cluster size. This
hyphothesis was based on the assumption of spherical and monodisperse
clusters, an assumption which is not strictly verified, as we have
already discussed, within the present system.  To quantify the
inter-cluster interactions, we adopt the following strategy.  We
consider a low enough $T$ where the cluster phase extends over a large
$\phi$ region. In each configuration, for each cluster we calculate
its center of mass coordinate. Then, independently of cluster size and
shape, we calculate the pair distribution function $g_{CM}^{clust}(r)$
between centers of mass of different clusters.  Results for $T=0.1$
and various studied $\phi$ are reported in Fig.~\ref{fig:grCM}.  It is
clear that the centers of clusters are found in liquid-like
configurations, with increasing correlations as $\phi$ increases up to
roughly $\phi=0.08$.  Next we compare the numerical
$g_{CM}^{clust}(r)$ with theoretical predictions obtained solving
numerically the Ornstein-Zernike (OZ) equation \cite{Lik01b} with the
hypernetted chain (HNC) closure. We checked that results are
independent on the chosen closure. As suggested in Ref.~\cite{Sci04a},
to model the cluster-cluster interaction, we select a pure Yukawa
potential (no hard-core), with the same screening length ($\xi=2.0$)
of the particle-particle repulsive interaction. We leave the amplitude
$A_{eff}$ of the Yukawa potential as the only fitting parameter.
Indeed, the number density of the clusters is read directly from the
simulation data.  The best-fit curves are reported in
Fig.~\ref{fig:grCM} and show a remarkable agreement with the data
extracted from the simulations.  The behavior of the cluster number
density and of the renormalized amplitude are reported in the insets.
$A_{eff}$ grows approximately linearly with $\phi$, starting from the
particle-particle value at $\phi\rightarrow 0$, while the number of
clusters progressively shrinks.

Hence, we can describe the system as composed essentially of
repulsively-interacting clusters up to $\phi=0.08$, with a renormalized
amplitude which results from the average of all amplitudes which
characterize clusters of different sizes and shapes.  This result is
central for the present paper, and confirms the conjecture put forward
in \cite{Sci04a}, which was then combined with the use of MCT to
provide evidence of the existence of a Wigner glass at low $\phi$.

The picture of interacting Yukawa clusters breaks for $\phi >
0.08$. The $g_{CM}^{clust}(r)$ can no longer be fitted with a
Yukawa analogue.  A careful look at the configurations shown in
Fig.~\ref{fig:clust} shows that, at this point, clusters have
completely lost their spherical-shape.  The radial distribution
function shows for $\phi \gtrsim 0.1$ a reversal of trend in the main
peak position and amplitude, which then persists at larger
$\phi$. Moreover, a small peak develops with increasing $\phi$ for $1
< r/\sigma < 2$, indicating the occurrence of some cluster branching
events. The Yukawa effective interaction naturally ceases to work in
this regime, and a competition between repulsion and increased
packing, favoring sometimes the formation of intercluster bonds,
emerges. This mechanism, which can be considered essentially absent
for $\phi \lesssim 0.08$, becomes important in this intermediate
regime $0.08 < \phi < 0.14$, and finally dominant above
percolation. These results highlight the mechanism by which a
crossover between a cluster phase and a percolating state is realized,
based on a change between repulsion-dominated clusters to a branching
regime of clusters due to increased packing.

\section{Dynamics: MSD and iso-diffusivity lines}

\begin{figure}[tbh]
\includegraphics[width=0.6\textwidth]{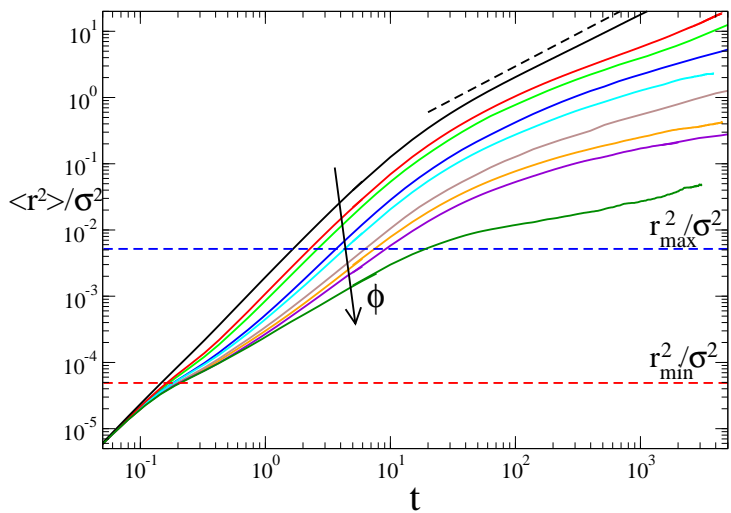}
\caption{MSD $<r^2>$ vs time for $T=0.1$ and various $\phi$. From top to
  bottom, $\phi=0.005, 0.02, 0.04, 0.08, 0.10, 0.125, 0.14, 0.16,
  0.20$. The dashed line indicates diffusive behaviour.
  Horizontal lines refer to the squared minimum $r_{min}^2$ and maximum
  bond distance $r_{max}^2$ (see text). }
\label{fig:msd}
\end{figure}

Next, we monitor the dynamics of the system, to find out whether and how we
approach dynamic arrest at low $T$, both in the cluster region and in
the percolating one. We start by calculating the particle mean squared
displacement (MSD) and reporting its behaviour in Fig.~\ref{fig:msd}
for $T=0.1$ and all studied $\phi$. We observe a gradual decrease of
the MSD with increasing $\phi$.  From the time dependence of the MSD
one can identify distinct regions. For very short times, one observes ballistic motion, followed by a 
slowing down at short times, which takes place approximately in correspondence with the length
scale of the attractive bond distance, defined as
$r_{min}/\sigma=(2^{1/100}-1)\simeq 0.007$, i.e. when two particles are
in the minimum of the potential well. 
This attractive localization by neighbouring particles is active 
for roughly a decade, even at small $\phi$,  in the cluster phase
region. It is quite remarkable to find particles rattling inside the
narrow bonds for such a long time.  For intermediate times, particles
are able to escape (on average) from the bond, as signalled by the
fact that they overcome the distance corresponding to the maximum bond
distance $r_{max}/\sigma=(r_b -1) \simeq 0.072$.   Hence, particles explore their
neighbourhoods, and a second slowing down emerges. This occurs on length scales that can exceed a particle diameter, depending on $\phi$. These results suggest that the 
motion of a single particle results from the  
sum of the brownian motion of the whole cluster and of  the intra-cluster motion.
We notice that only the smallest values of $\phi$ are able to recover a pure diffusive
regime at this temperature. For larger $\phi$
a sub-diffusive growth of the MSD is found in the time window which can be numerically studied. The apparent exponent regulating the
subdiffusivity is found to decrease with $\phi$, as expected for an MSD 
approaching a flat plateau. The onset of subdiffusive behaviour
signals that arrest is close-by, and located at slightly lower $T$,
both for the cluster phase region and for the percolating one. In both
cases, particles are trapped at first by the attractive bonds with
neighbouring particles and secondly by a larger localization
length. In the case of a cluster phase, such length provides the
distance where clusters can rattle (see Movies provided as supplementary information\cite{movies}). For the
percolating states, the localization length does not decrease much and
the arrested state can be identified as a gel\cite{Zac07a}.
Interestingly, the transition between the two arrested states is
continuous from the MSD point of view.

\begin{figure}[tbh]
\includegraphics[width=0.6\textwidth]{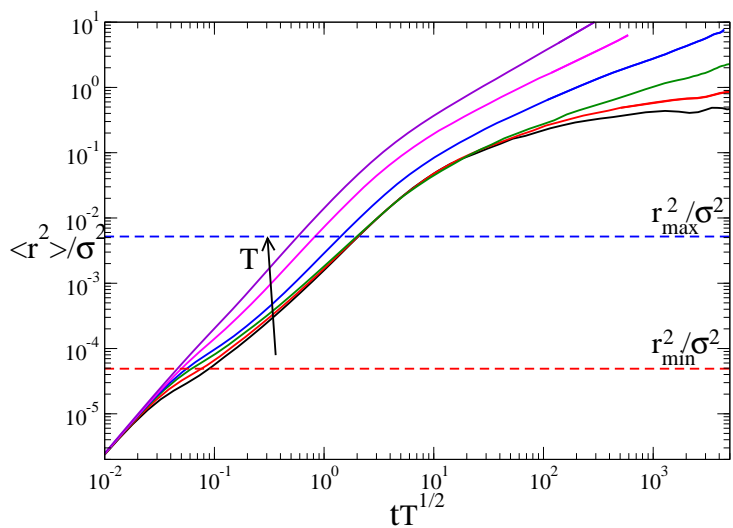}
\caption{MSD  $<r^2>$ vs time for $\phi=0.125$ and various $T$. From top to
  bottom, $T=0.2, 0.15, 0.125, 0.10, 0.07, 0.05$.  Horizontal lines
  refer to the squared minimum $r_{min}^2$ and maximum bond distance
  $r_{max}^2$ (see text).  }
\label{fig:msdT}
\end{figure}

To further prove that arrest takes place also in the non-percolating
regime, we show in Fig.~\ref{fig:msdT} the MSD for $\phi=0.125$
and various studied $T$. We remark that this is an isochore along which the system
never percolates, at all studied $T$, but is the closest, among the studied $\phi$, to the
percolation transition. Similar results are found at lower $\phi$. At very low $T$, the MSD approaches a
flat behaviour at long times, so that true arrest occurs. The low-$T$
curves are calculated within a time-window where the energy
slowly drifts, and very slow aging effects are thus present. These effects can only
slow further the relaxation, so that the reported curves can be
considered  an overestimation of the true MSDs.  The
localization length for the arrested state at this $\phi$ can be
estimated around $\approx 0.8\sigma$. The effect of temperature is also visible in the
amplitude of the vibration within the bond, which decreases with $T$, as seen in the 
height of the inflection around $r_{min}$. 

We note in passing that, in principle, we can also calculate the MSD
of clusters (not shown), by monitoring their center of mass trajectory. Due to the
fact that clusters undergo breaking and reforming events during the
simulation time, statistics is poor. However, in all studied cases,
the average MSD of clusters center of mass is always found below that
of a single particle. This provides a hint of the fact that, in the explored time window,
most of the delocalization is provided by rotational motion of the isolated clusters (see also Movies provided in the supplementary information\cite{movies}).

\begin{figure}[tbh]
\includegraphics[width=0.6\textwidth]{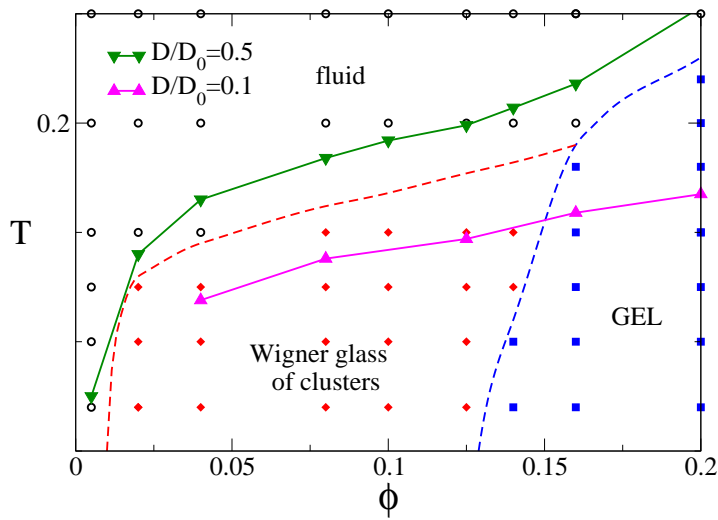}
\caption{Phase diagram, partially redrawn from
  Fig.~\protect\ref{fig:phase}, with added iso-diffusivity lines, for
  two selected values of $D/D_0$.  }
\label{fig:phase-isoD}
\end{figure}

From the long-time behaviour of the MSD, we can extract the
self-diffusion coefficient, defined as $D=\lim_{t\rightarrow\infty}
\langle r^2\rangle/6t$, for those state points where a diffusive
long-time regime can be clearly identified. We can draw
iso-diffusivity lines, i.e. loci in the phase diagram with constant
diffusion coefficient, expressed as a fraction of the bare diffusion
coefficient $D_0$.  We plot some iso-diffusivity lines in
Fig.~\ref{fig:phase-isoD}, together with the phase diagram reported
above.  We clearly find that the iso-$D$ lines follow at small $\phi$ the
shape of the cluster phase boundary, while at larger $\phi$ they follow the
percolation boundary. From
previous studies\cite{Zac02a,Fof03a,Kum05a,Zac06JCP}, we know that
the shape of iso-diffusivity lines does not change much with
approaching distance to the $D=0$-line, which can be identified with
the ideal arrest transition. Hence, if one could extrapolate, the
arrest line would be somehow parallel to the iso-$D$ lines, signaling
that arrest is mainly temperature-driven. Again, we notice the
continuous shape of the iso-diffusivity lines across the percolation
transition. This allows us to identify low-$T$ states as arrested, or
approaching arrest at low $T$, independently of the presence of a
percolating network, the latter condition being discriminant for
determining the nature of the two arrested states.  Hence, we can
conclude that also the non-percolating states undergo dynamical arrest
at low $T$ in full analogy with the percolating ones. Thanks to the
identification of the dominant cluster-cluster interactions discussed
above, arrest at low $\phi$ can  therefore be interpreted as a Wigner glass of
clusters, and it is clearly distinct from arrest driven by the formation of a spanning cluster (gel).

 Due to the presence of subdiffusion and of aging effects at low $T$,
 we can only probe a limited window  in $D/D_0$. 
 The behaviour of $D$ vs $T$ (not shown) displays a rapid decrease below $T=0.20$,
 then followed by a slower (non-Arrhenius) decrease for $T\lesssim 0.125$.  Due to the
 limited available data and the difficulty to extract $D$ for the
 low-$T$ states, our analysis can not be detailed. However, the
 apparent non-Arrhenius behaviour might indicate the presence of a finite-$T$
 arrest line, at least for the larger densities, and in particular in
 the gel regime. This is different from what was reported for other
 gel-forming systems, based on patchy or limited-valence interactions,
 where Arrhenius behaviour has been observed\cite{Zaccagel,Bia06a,DelG07}, 
 compatible with an ideal gel state occuring only at $T\rightarrow
 0$\cite{Zac07a,genova}.  Indeed, in the patchy particle models, the relaxation and approach to
 a gel state are essentially controlled by the
 single-bond lifetime\cite{Zac06JCP} and bonds were shown to be independent.
 For the present case, the interplay between different types of interactions, and in particular
 their long-range nature, might be responsible for the fact that
 collective bond rearrangements are needed in order to restructure the system. 
 
\section{Conclusions}
In this manuscript, we have reported extensive Brownian Dynamics
simulations of a model potential suitable to describe the interactions between charged colloidal
particles in apolar solvent, in the presence of an additional short-range depletion
attraction.  The long-range nature of the repulsive screened Coulomb
interactions is taken into account numerically by performing Ewald
summation.  We have studied a wide region of the $(\phi, T)$ plane,
covering the evolution of the system from a fluid of monomers to 
a percolating network. In between these two limits, a
region of stable finite-size clusters is found. These clusters are
spherical at low $\phi$, acquiring more and more elongated shape with
increasing $\phi$, due to the role of cluster-cluster interactions.

We have provided evidence that particles inside clusters can acquire
local order, but no long-range order, both inside clusters and
among different clusters, is present. Most importantly, we have been
able to quantify the cluster-cluster interactions in terms of an
effective Yukawa potential, with the same screening length as for
particle-particle interactions, and a renormalized amplitude,
confirming a previously hyphothesized scenario \cite{Sci04a}. This
happens independently of the clusters polydispersity in size and
shape. However, when $\phi$ increases, the situation changes, and
cluster-cluster interactions become more subtle, due to the increasing
number of branching events and to the elongation of the
clusters.

At low temperatures, we find evidence of dynamic arrest, 
by monitoring the MSD and the particles  self-diffusion coefficient. The
arrest mechanism appears to be continuous across all studied $\phi$
and  is mostly driven  by temperature. However, the distinct nature of the
two different low-$T$ states, namely  finite-size clusters or spanning network, 
allows us to unambiguously identify the presence of two different non-ergodic states in these
colloidal systems.  A Wigner glass of clusters exists at
low/intermediate $\phi$, stabilized by the renormalized Yukawa
cluster-cluster interactions discussed above, while a gel state,
stabilized by the presence of a percolating, long-lived network takes
place at larger $\phi$.  It will be interesting in future studies to
compute the viscoelastic response of these two different non-ergodic
systems, that should manifest extremely different rheological
properties. The accurate knowledge of the phase diagram, provided in
this work, will allow us to choose the best conditions where such
response is measurable and to compare it to those of other
gel or glass-forming systems, as well as to experimental results.

\section{Acknowledgements}
We thank S. Mossa for useful discussions on the isolated cluster ground state properties. EZ
wishes to thank R. Hidalgo-Alvarez and the Fluid and Biocolloids Group at the
University of Granada for kind hospitality during her stay in Granada, when part of this work was performed. JCFT thanks Junta de Andalucia for financial support during his stay in Rome. We acknowledge support from Proyecto de Excelencia de la Junta de Andalucia P05-FQM-0392, the Marie Curie Network on Dynamical Arrest of Soft Matter and Colloids MRTNCT-2003-504712 and NoE SoftComp NMP3-CT-2004-502235.

\bibliographystyle{./jpc}
\bibliography{./articoli,./altra,./biblio_patchy,./star-star}

\end{document}